\documentclass[useAMS,usenatbib,usegraphicx]{mn2e}
\usepackage{natbib}
\usepackage{color}

\newcommand\bfk{\mbox{$\bf k$}}

\newcommand{\apjl}{ApJ}
\newcommand{\apjs}{ApJS}
\newcommand{\aap}{AAP}

\title{Cosmology without cosmic variance}
\author[Bernstein \& Cai]{Gary M. Bernstein\thanks{garyb@physics.upenn.edu } \& Yan-Chuan Cai \\
 Dept. of Physics and Astronomy, University of Pennsylvania}

\begin{document}
\maketitle

\begin{abstract}
We examine the improvements in constraints on the linear growth factor $G$ and its derivative $f=d\ln G / d\ln a$ that are available from the combination of a large-scale galaxy redshift survey with a weak gravitational lensing survey of background sources. In the linear perturbation theory limit, the bias-modulation method of McDonald \& Seljak allows one to distinguish the real-space galaxy clustering from the peculiar velocity signal in each Fourier mode. The ratio of lensing signal to galaxy clustering in transverse modes yields the bias factor $b$ of each galaxy subset (as per Pen 2004), hence calibrating the conversion from galaxy real-space density to matter density in every mode.  In combination these techniques permit measure of the growth rate $f$ in each Fourier mode.  In principle this yields a measure of the growth rate that is free of sample variance, {\it i.e.} the uncertainty in $f$ can be reduced without bound by increasing the number of redshifts obtained within a finite survey volume.  In practice, the gain from the absence of sample variance is bounded by the limited range of bias modulation among dark-matter halos. Nonetheless, the addition of background weak lensing data to a redshift survey increases information on $G$ and $f$ by an amount equivalent to a 10-fold increase in the volume of a standard redshift-space distortion measurement---if the lensing signal can be measured to sub-percent accuracy. This argues that a combined lensing and redshift survey over a common low-redshift volume of the Universe is a more powerful test of general relativity than an isolated redshift survey over larger volume at high redshift, especially as surveys begin to cover most of the available sky.  An example case is that a survey of $\approx10^6$ halo redshifts over half the sky in the redshift range $z=0.5\pm0.05$ can determine the growth exponent $\gamma$ for the model $f=\Omega_m^\gamma$ to an accuracy of $\pm0.015$, using only modes with $k<0.1h\,{\rm Mpc}^{-1}$, but only if a weak lensing survey is conducted in concert.
\end{abstract} 


\section{Introduction}
The growth of large-scale structure in the Universe is a competition between gravitational attraction and the expansion of the Universe. In the linear-perturbation-theory limit (LPT), each Fourier mode evolves independently, and the mass density fluctuation amplitude at a comoving wavenumber \bfk\ is $\delta(\bfk, a) = G(a) \delta(\bfk, a_0)$, with $a$ the scale factor. Knowledge of the linear growth function $G(a)$ would, in standard general relativity, reveal the expansion history of the Universe, including any effects of ``dark energy.'' Given prior knowledge of the expansion history $a(t)$, the measure of $G(a)$ tests the time and scale dependence of any deviation from General Relativity.\footnote{We will ignore possible scale dependence of $G(a,k)$, that could arise from modifications to General Relativity, but this is merely a notational convenience: all the results in this paper apply equally well to a scale-dependent $G$ or $f$.}  The surprising observation that the growth $a(t)$ is accelerating in recent epoch strongly motivates tests of the accuracy of General Relativity at the largest observable physical scales, preferably to accuracy $<1\%$. 

Since the growth of large-scale structures is unmeasurably small over the brief history of human astronomical observations, we cannot directly track the growth in a given mode. Growth measures must adopt one of two alternative strategies. The first is to compare fluctuations at the same $k$ at two different epochs on the light cone, invoking the Cosmological Principle to permit comparison of structure at distinct locations.  The comparison is therefore limited by the sample variance from the finite number $N_m$ of mode amplitudes that are drawn from the distribution and observable at each epoch: the power $P$ is measured to  accuracy $\sigma_P / P \ge \sqrt{2/N_m}$ regardless of the noise level in the observed fluctuation field.  If the theoretical prediction for $P$ is sufficiently accurate for $k<k_{\rm max}$, then the number of modes available to a survey of volume $V$ is $N_m=Vk^3_{\rm max}/6\pi^2$. 

The sample variance from the finite observable volume of the Universe is often called {\em cosmic variance} and places a fundamental limit to the measure of growth by power-spectrum comparison in a chosen $k$ range.  The sample variance limits become particularly acute at low redshifts, because the available volume scales as $z^2\Delta z$ and because non-linear growth of structure reduces the $k_{\rm max}$ at which accurate theory is possible.  This is unfortunate since the manifestations of dark energy or modified gravity are thought to be the largest at the present epoch.

The second strategy for measuring growth is to invoke the continuity equation to link the growth rate to the velocity amplitude in each mode. If each mode evolves independently, and the mass is conserved with a single-valued peculiar velocity ${\bf v}$ at each location, then the continuity equation ${\partial \delta \over \partial t}+\nabla \cdot(1+\delta){\bf v}=0$ becomes, to first order in the density perturbation, $\nabla \cdot {\bf v} = -{\partial \delta \over \partial t}$.
In Fourier space, it yields
\begin{equation}
{\bf v}(\bfk) = f H\delta(\bfk) {i\bfk \over k^2},
\end{equation}
where $f \equiv { \partial \ln G \over \partial \ln a}$ is the growth rate and $H$ is the Hubble parameter at that time.
In real space or Fourier space, the velocity field is proportional to the mass density field, with the constant of proportionality $f$ delivering the desired information on the (differential) growth of structure.  Since the LPT velocity field is irrotational, it suffices to map one component of the velocity field.
Massive observational efforts were undertaken in the 1980's and 1990's to map the density and (radial) velocity fields of galaxies in the nearby ($<100h^{-1}$~Mpc) Universe.  These were stymied by (among other issues) two difficulties: first, the observed radial velocity of a galaxy at distance $r$ is $v_{\rm obs} = {\bf v}\cdot{\bf r}/|r| + Hr$, so these surveys required distance indicators for each galaxy to determine $r$ and the (radial) peculiar velocity $v$ indendently, introducing large statistical and systematic errors.  Second, even with perfect distance indicators, the density field would be derived from the galaxy field, so the ratio of velocity to density will yield $f/b$, not $f$. Inferences on the growth rate $f$ can be only as accurate as prior knowledge of the galaxy bias factor $b$.

Current proposals for measuring $f$ avoid the need for distance indicators---thus extending the measurements to larger volumes and higher redshifts---by analyzing the directional dependence of the redshift-space galaxy power spectrum $P^s(\bfk)$.  If we survey a tracer population that follows the velocity field of the matter, \citet{Kaiser} shows that the LPT value of its Fourier fluctuation amplitude $\delta^s$ in redshift space is related to the real-space amplitude $\delta^r$ of the tracers and $\delta$ of the mass by
\begin{equation}
\delta^s(\bfk) = \delta^r(\bfk) + f\mu^2 \delta(\bfk).
\label{deltas}
\end{equation}
Here $\mu$ is the cosine of the angle between $\bfk$ and the line of sight (taking the plane-parallel approximation). The standard simple approach is to next assume that the tracer density has some bias $b$ with respect to the matter, so that $\delta^r =b\delta$, but faithfully traces the matter velocity field on large scales, in which case the redshift-space power spectrum becomes
\begin{eqnarray}
P^s(\bfk) & = & (b^2 + 2bf\mu^2 + f^2 \mu^4) P(k) \nonumber \\
 & = & f^2P(k)\left[ (b/f)^2 + 2(b/f)\mu^2 + \mu^4\right].
\label{ps1}
\end{eqnarray}
Measurement of $P_s(k,\mu)$ again cannot constrain $f$ without prior knowledge of $b$, however the combination $f^2P$ can be constrained.  Current state of the art is demonstrated by the $\approx 20\%$ measures of $f^2P$ in bins of width $\Delta z=0.2$ by \citep{Blake}.  We have $f^2P=f^2G^2P_0$, where $P_0$ is an ``initial'' power spectrum at an early epoch, e.g. recombination, and $G$ is the linear growth since that epoch.  Relying on the CMB to determine $P_0$, we can constrain the growth quantity $fG$ by comparing the power in distinct modes (different $\mu$ instead of different epochs this time), which again leads to a fundamental sample-variance floor.  Note also that any uncertainty in $P_0$, for example from uncertainty in the reionization history, propagates into an error of half the size in $fG$.  
The regression against $\mu$ necessary to extract $fG$ amplifies the sample variance significantly.  A Fisher analysis of the standard RSD analysis in the sample-variance limit yields
$\sigma_{\ln fG} \approx \sqrt{21/N_m}$ for the case $b/f=1.4$ (with the prefactor becoming even less favorable for larger $b/f$).  

\begin{figure}
\resizebox{\hsize}{!}{
\includegraphics[angle=0]{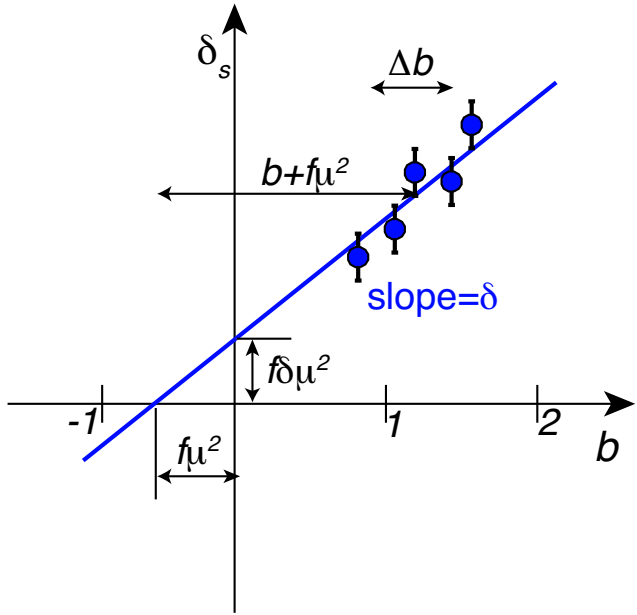}}
\caption[]{
Schematic illustration of the McDonald-Seljak technique: in a chosen Fourier mode with real-space mass density fluctuation $\delta$, we observe the redshift-space amplitude $\delta^s_i$ of galaxies with different biases $b_i$.  If the Kaiser formula (\ref{deltasi}) holds, then linear regression of the $\delta^s_i$ data points against bias will yield the $y$-intercept value  $f\mu^2\delta$ and the $x$-intercept value $f\mu^2$.  The former is one sample from a Gaussian with variance $f^2\mu^4P$.  The latter gives $f$ without sample variance.  However the uncertainty in $f$ from this mode's data is amplified if $\mu\ll1$, or if the range $\Delta b$ of galaxy biases is small compared to the typical $b+f\mu^2$ value.
}
\label{msfig}
\end{figure}

\citet{MS}[MS] propose an important improvement to the peculiar-velocity measurement. Divide the galaxy sample into subsets with differing bias factors $b_i$, and assume that the overdensity of each bin in redshift space follows the Kaiser form:
\begin{equation}
\delta^s_i = (b_i + f \mu^2) \delta + \epsilon_i,
\label{deltasi}
\end{equation}
with $\langle \epsilon_i \epsilon_j \rangle = \delta^K_{ij} / n_i$ describing a diagonal stochasticity matrix.  For the usual assumption of Poisson sampling, $n_i$ is the space density of the objects in bin $i$.  In the limit of $1/n_i\rightarrow 0$, MS note that observation of the $\delta^s_i$ in a {\em single mode} at $\mu=0$ will yield the values of $b_i$ up to an unknown normalization $\bar b$.  Then a subsequent noiseless observation of another single mode at $\mu\ne 0$ will allow one to regress $\delta_i^s$ against $b$ to determine both $f\mu^2\delta$ and $\bar b \delta$, as illustrated in Figure~\ref{msfig}.  Since $\mu$ is known, in the limit of low stochasticity this method produces
\begin{enumerate}
\item a measure of $f^2P=f^2G^2P_0$ yielding uncertainty $\sigma_{\ln fG} = 1/\sqrt{2N_m}$, assuming known $P_0$.  This is the sample-variance limit if we could simply view the ``naked'' velocity field distinct from the real-space clustering, and is a factor of $\sim 40$ improvement in variance over the standard  (RSD) analysis, which was degraded by the need to regress $P^s$ against $\mu$;
\item a measure of $f / \bar b$ from taking the ratio of velocity to real-space density amplitudes in {\em each mode}.  Since the measure can be made mode-by-mode, it is {\em not limited by sample variance}, i.e. can in theory reach unlimited precision from a finite set of modes.
\end{enumerate}
When the $n_i$ are finite, these gains are of course ameliorated.  Numerical results are published for constraints on $fG$ expected from values of $n_i$ and $b_i$ that are arbitrarily assigned in 2 bins (MS); or for halos split into 2 bins \citep{GM10}; or for potential observed samples split into 4 bins \citep{WSP09}.  In this paper we derive the optimal results attainable for the MS method using any number of bins of halos.

The sample-variance-free measure of $f/\bar b$ is not useful for cosmological constraint without further information on $\bar b$.  Cross-correlation of galaxy redshift surveys with weak gravitational lensing shear fields---which are a direct measure of the true mass overdensity $\delta$---yields a sample-variance-free measure of $b$ \citep{Pen04}, and MS comment that combining this technique with the multiple-tracer RSD measure could yield improved cosmological constraints.  The main result of this paper is to quantify the gain in growth-of-structure constraints from this combination of galaxy redshift surveys with weak lensing surveys.  This combination is attractive for several reasons:
\begin{enumerate}
\item The combination of bias-modulated RSD data with lensing data yields an estimate of $f$ that is free of sample variance, meaning arbitrarily good accuracy with finite survey volume.  This is very attractive for measuring growth on very large scales at low $z$, where Fourier modes are scarce.
\item This combination measures $f$ directly, breaking degeneracies with $b$ or $P$.  Consequently the cosmological model can be constrained without being subject to uncertainties in the CMB power spectrum normalization due to reionization complexities, potentially even exceeding the cosmic variance level of the CMB measurement.
\item Breaking the degeneracy between $f$ and $P$ enables a consistency check between the growth function $G$ and its derivative $f$ over the observed redshift range.  This allows some of the theoretical assumptions to be checked, for example the assumption that the continuity equation for mass can be used to infer the velocity field for halos or galaxies.
\end{enumerate}

We first give a qualitative discussion of the types of measurements that are free of sample variance, then we provide the analytic formalism and derive the scaling of the errors on $f$ and $P$ in the case of independently-determined biases.   We do not limit our analysis to 2 bins of galaxy bias.  We next provide a numerical analysis of the growth constraints available in a candidate redshift survey of all halos above mass $M_{\rm min}$ in a volume centered at $z=0.5$.  This is a redshift at which the standard RSD analysis is becoming very constrained by cosmic variance. 
Unless noted otherwise, we assume a fiducial flat $\Lambda$CDM cosmology with the following parameters: $\Omega_m$=0.27, $\Omega_{\Lambda}=0.73$, $\Omega_b=0.046$, $\sigma_8=0.9$, $n_s=1$, $H_0=72$.

\section{Sample-variance-free measurements}
Consider a random variable $\delta$ that is drawn from a zero-mean Gaussian distribution with variance $P$.  If we draw $\{\delta_1, \delta_2,\ldots, \delta_{N_s}\}$ from the distribution, and execute a measurement process on each sample that yields $\hat\delta_i = \delta_i + \epsilon_i$ due to some measurement noise $\epsilon_i$, then how well can we determine the power $P$?  If the noise power ${\cal E}=\langle \epsilon^2 \rangle$ is known, then the uncertainty on $P$ satisfies
\begin{equation}
\sigma^2_{\ln P} = {2 \over N_s} \left( 1 + {\cal E}/P \right)^2.
\label{siglnp}
\end{equation}
The constant term inside the parentheses sets a sample-variance limit that cannot be reduced by better measurements, only by obtaining more samples.  If $\delta$ is the mass density or gravitationl potential fluctuation in a Fourier mode, then $N_s$ is the number of independent modes in the surveyed volume, which we call $N_m$:
\begin{equation}
N_m = {V \over (2\pi)^3} {4 \pi k_{\rm max}^3 \over 3}.
\end{equation}

Next consider the case where two tracers of the underlying random variable $\delta$ are available.  Two measurements, $\hat\delta_i^A = \delta_i + \epsilon_i^A$ and $\hat\delta_i^B = b\delta_i + \epsilon_i^B$, are available for sample $i$.  Tracer B is biased by some factor $b$ with respect to the underlying field $\delta$, and measurement errors are uncorrelated.  The covariance matrix between $\hat\delta^A$ and $\hat\delta^B$ is
\begin{equation}
C_{AB} = \left( \begin{array}{cc}
P + {\cal E}^A & bP \\
bP & b^2P + {\cal E}^B
\end{array} \right).
\end{equation}
If we assume that ${\cal E}^A$ is known but that $P$, $b$, and ${\cal E}^B$ are free parameters, we can derive the covariance matrix for our estimates of these parameters using the Fisher matrix formalism.  The covariance matrix for these three parameters is, at best, the inverse of the Fisher matrix $F$ having components \citep{TTH}
\begin{equation}
F_{ij} = \frac{1}{2}{\rm Tr}\left(C_{AB}^{-1} C_{AB,i} C_{AB}^{-1} C_{AB,j}\right).
\end{equation}
The uncertainty in $\ln P$ in this two-tracer case remains the same as (\ref{siglnp}), and has the same sample-variance limit $\sigma_{\ln P}\ge\sqrt{2/N_s}$.  The uncertainty on the bias parameter $b$ (after marginalization over $P$ and ${\cal E}^B$) is
\begin{equation}
\sigma^2_{\ln b} = {1 \over N_s} \left[  \frac{{\cal E}^A}{P} + \frac{{\cal E}^B}{b^2P} + O({\cal E}/P)^2 \right].
\label{sigb}
\end{equation}
Note that the leading order term in ${\cal E}/P$ is linear: the sample-variance term is missing, and a decrease in measurement error reduces uncertainty in $b$ without bound.  The fundamental difference between $b$ and $P$ is that the former can be measured accurately from a single pair of samples $\hat\delta^A$ and $\hat\delta^B$ whereas the latter is a population property.  An intuitive result is that, with no measurement noise, $b=\hat\delta^B / \hat\delta^A$ can be measured perfectly as the slope of the locus of measurements in the $\hat\delta^A-\hat\delta^B$ plane.

We take advantage in this paper of three circumstances in which cosmological information can be extracted from the {\em relative} signals of distinct tracers of the mass density fluctuation field.
\subsection{Weak lensing convergence vs galaxy distribution}
For fluctuation modes transverse to the line of sight, gravitational lensing imparts a shear field on background galaxies that can be measured with weak lensing techniques.  The associated lensing convergence amplitude $\kappa$ is equal to the matter fluctuation $\delta$, modulo some distance factors, which we will assume are known. The second mass tracer is the galaxy density map, which has an unknown bias factor $b$ with respect to matter.  \citet{Pen04} and \citet{BJ04} both note that the proportionality constant between lensing and galaxy signals can be measured without sample variance.  For the lensing signal, the measurement error ${\cal E}^A$ is from shape noise, and at large scales ($\ell\approx 100$) the ratio ${\cal E}/P$ for ``cosmic shear'' power is $<0.1$ in each mode \citep{TakadaJain}.   For the galaxy tracer, the measurement error ${\cal E}^B$ arises from stochasticity in the galaxy distribution, often assumed to be Poisson noise, in which case ${\cal E}^B/b^2P=(nb^2P)^{-1}$, where $n$ is the space density of galaxies, and this can be quite small.  \citet{CBS11} show that an optimally-weighted survey of halos can reproduce the mass density with very low stochasticity---{\it e.g.} an optimally weighted survey of halos with $M>10^{13.5}h^{-1}M_\odot$ can obtain ${\cal E}/b^2P<0.1$.  Equation (\ref{sigb}) then suggests that the comparison of lensing and galaxy information in the $\approx 10^4$ transverse modes with $\ell \le 100$ should yield bias uncertainties $\sigma_{\ln b}<0.01.$


We defer to future work the examination of the complexities in constraint of galaxy bias by this method, such as the simultaneous solution for bias and distance factors, and the superposition of many galaxy redshifts along a given lensing line of sight.  In this work, we will simply assume that large-scale galaxy biases---or at least the bias of a chosen weighted combination of halos---can be derived by the lensing$+$galaxy surveys to potentially sub-percent accuracy, limited by signal-to-noise rather than by sample variance.

\subsection{Real-space clustering for different bias bins}
Consider next a redshift survey that has targets divided into bins expected have different bias, as per MS.  We will consider, optimistically, that redshift targets are dark-matter halos of known mass, and that we bin the targets by halo mass.
For a transverse mode ($\mu=0$), the redshift-space survey has no velocity contribution [Equation~(\ref{deltasi})].  As noted by MS, the ratio $b_j/b_i$ of biases of two different mass bins $i$ and $j$ can be measured without sample variance, since this is the ratio of two tracers of the same underlying field $\delta$.   Hence the biases of all halo bins can be determined, without sample variance, up to an overall scaling factor $\bar b$.  The addition of lensing data allows a high-precision determination of $\bar b$ as described above, again without sample variance, so the combination of both techniques will yield biases of all halo-mass bins.

\subsection{Velocity field vs density field}

 MS offer a third means of extracting cosmological information from a ratio of tracers. With the redshift-survey target galaxies divided into bins of differing bias, it is possible to distinguish the two terms $b_i\delta$ and $f\mu^2\delta$ that contribute to the amplitude of the redshift-space amplitude $\delta^s_i$ in every mode.  The ratio of these two components is $f\mu^2 / b_i$, another quantity that can therefore be measured without a sample-variance limitation.  If the $b_i$ are known, $f$ is obtained.

We see therefore, that combining a lensing survey with a redshift survey over the same volume provides a means to measure the ratio of the velocity signal to the {\em mass} overdensity in every (non-transverse) mode.  This ratio gives $f$, without any fundamental limit from sample variance, without the need to compare to a power spectrum at a reference epoch, and without any dependence on bias.

\section{Forecast methods}
We construct a Fisher information matrix for the constraint of growth-related parameters by a canonical redshift survey of Fourier modes with $k<k_{\rm max}$ in volume $V$ at redshift $z$.  The redshift-space structure is taken to follow the Kaiser formula (\ref{deltasi}), and the real-space matter fluctuations are taken to obey $\langle \delta^2({\bf k})\rangle = G^2P_0(k)$ with no correlation between modes.  The shape $P_0(k)$ of the power spectrum is taken as known, while the amplitude $G$, {\it i.e.} the growth function, is a parameter of interest.  The $\delta^s_i$ fluctuations in a single Fourier mode of the survey volume are distributed, in LPT, by a multivariate zero-mean Gaussian with covariance matrix
\begin{eqnarray}
\label{cij}
C_{ij} & = & {\rm Cov}(\delta^s_i,\delta^s_j) = (b_i+f\mu^2)(b_j+f\mu^2) G^2P_0 + {\cal E}_{ij} \\
{\cal E}_{ij} & \equiv & \langle \epsilon_i \epsilon_j \rangle
\end{eqnarray}
The model has free parameters (which we index by Greek letters) $\theta_\alpha \in \{f, G, b_1, b_2, \ldots, b_N\}$.  In the model where the spectroscopic targets are halos that sample the mass field by a pure Poisson process, we will have the stochasticity matrix
${\cal E} = {\rm diag}({\cal E}_i) \equiv {\rm diag}(1/n_i)$, where $n_i$ is the space density of targets in bin $i$.  \citet{Hamaus10} and \citet{CBS11} 
demonstrate that the simple diagonal Poisson-noise formula for ${\cal E}$ does not fully describe the stochasticity of halos in $N$-body simulations.  We have verified that one can treat the {\em diagonal} elements ${\cal E}_i$ as free parameters of the model, with insignificant degradation of the growth measurement, so this part of the Poisson assumption is not critical.  The success of the analysis described herein does, however, depend crucially on the assumption that the {\em off-diagonal} elements of ${\cal E}$ vanish, so future work will need to quantify the corrections needed for non-diagonal stochasticity.  

The Fisher matrix for information from a single mode follows the standard form for multivariate, zero-mean Gaussian data \citep{TTH}:
\begin{equation}
F_{\alpha\beta} = \frac{1}{2} {\rm Tr}\left[ C^{-1} \frac{\partial C}{\partial\theta_\alpha} C^{-1} \frac{\partial C}{\partial\theta_\beta}\right].
\end{equation}
The total Fisher matrix is then the sum over all independent ${\bf k}$ modes in the survey volume.

The total Fisher matrix is then used to predict the covariance matrix of $\ln f$ and $\ln G$ after marginalization over all other parameters, namely the biases.  Marginalization and mapping to new parameters are done with standard techniques, {\it e.g.} as summarized by \citet{FoMSWG}.

\subsection{Cases}
We will forecast uncertainties on the growth parameters for the following types of experiments:
\begin{itemize}
\item {\bf Standard RSD:} All halos with mass $M>M_{\rm min}$ in the survey volume have redshift measurements and are combined into a single sample.  We marginalize over the bias of this combined sample.  We know that there will be a degeneracy between $\ln f$ and $\ln G$ since only the quantity $f^2P$ can be constrained in this experiment.  We will report the uncertainty on the quantity $\ln fG$.  As noted in the introduction, there is a sample variance limitation $\sigma_{\ln fG}\ge \sqrt{21/N_s}$ for $b/f=1.4$.
\item{\bf MS:} The same as Standard RSD, except that halos are binned by mass and the analysis suggested by MS is performed.  Biases of every mass bin are free parameters.  We typically use 10--20 mass bins, which will produce results indistinguishable from the ideal limit of an analysis without information loss due to binning.  Again $\ln f$ and $\ln G$ are degenerate without prior knowledge of biases, so we report uncertainties on $\ln fG$.  The sample variance limitation is $\sigma_{\ln fG}\ge1/\sqrt{2N_m}$.
\item{\bf Fixed Biases:} The same as the MS analysis except this time we assume that biases of all halo mass bins are known perfectly through combination with a weak lensing survey of the same sky area.  In this case we can simply strike the bias-related elements from the Fisher matrix.  The degeneracy between $f$ and $G$ is broken so we can report two uncertainties $\sigma_{\ln f}$ and $\sigma_{\ln G}$.  Constraint on alternative growth scenarios must account for possible covariance between $\ln f$ and $\ln G$ in the measurement as well.  The sample-variance limit for $\sigma_{\ln G}$ is $1/\sqrt{2N_m}$, the same as for $fG$ in the MS method.
\item{\bf Weighted Bias:} A more realistic scenario than the Fixed Bias case is that the weak lensing measurements provide an estimate of the bias of a single weighted combination of the halos.  This would be found by cross-correlation of the lensing convergence with a weighted map of halos.  We assume that the bias of this weighted halo population is determined to a fractional accuracy of $\sigma_{\ln b}$. The effect on the Fisher matrix is to add $w_i w_j b_i b_j/ \sigma^2_{\ln b}$ to the element of the Fisher matrix linking bias $b_i$ with bias $b_j$. Here $w_i$ is the weight on halos in bin $i$, normalized so that $\sum w_i=1.$.  We marginalize over all the biases after adding this prior on the weighted bias. We assume that mass weighting would be used, as a simple approximation to the minimum-stochasticity weights derived by \citet{Hamaus10} and \citet{CBS11}. 
The weighted-bias-measurement experiment breaks the degeneracy between $f$ and $G$.
\end{itemize}

\subsection{Algebraic Results}
Algebraic expressions for the Fisher and covariance matrices in these cases are available, because the matrix $C$ is invertible via the Sherman-Morrison formula.  Defining a vector ${\bf u}$ with $u_i = b_i + f\mu^2$, we have
\begin{eqnarray}
C & = & {\cal E} + {\bf u} G^2P_0 {\bf u}^T \\
\Rightarrow \quad C^{-1} & = & {\cal E}^{-1}- \frac{G^2P_0{\cal E}^{-1} {\bf u} {\bf u}^T {\cal E}^{-1}}{1+G^2P_0{\bf u}^T{\cal E}^{-1}{\bf u}}.
\end{eqnarray}
In the Poisson case, ${\cal E}^{-1}={\rm diag}(n_i)$ and the Fisher-matrix elements all can be expressed in terms of number-weighted sums over the halo population.  These results are stable under change in bin size as long as the bias values do not vary widely within any given bin.
Marginalizing over bias and summing over the distribution of $\mu$ for all the modes leads to rather lengthy, opaque expressions for the final uncertainties.  However in the case of a single mode at fixed $\mu$ with known biases, the uncertainties in $G$ and $f$ can be derived in an illuminating form.  If we define $n=\sum n_i$ as the total space density of halos surveyed, we obtain
\begin{eqnarray}
\sigma^2_f & = & \frac{ 1 + nP \left\langle(b+f\mu^2)^2\right\rangle}{\mu^4 (nP)^2 {\rm Var}(b)} \\
& \rightarrow & \frac{1}{\mu^4 nP}\frac{\left\langle(b+f\mu^2)^2\right\rangle}{{\rm Var}(b)} \qquad (nP\gg 1), \\
\frac{\sigma^2_G}{G^2} & \rightarrow & \frac{1}{2} \qquad (nP\gg 1). 
\label{sigmaf}
\end{eqnarray}
The averages and variance in this expression are taken over the targets of the redshift survey.

This scaling is understood from examination of Figure~\ref{msfig}, illustrating the MS method.  The quantity of interest, $f\mu^2$, is equal to the $b$-intercept of the linear regression of the fluctuation amplitudes $\delta_i$ against the bias values $b_i$.  
The main uncertainty $\Delta(f\mu^2)$ in the location of this intercept come from statistical fluctuations $\Delta m$ in the fitted slope $m$ of $\delta_i$ vs $b_i$:
\begin{equation}
\Delta(f\mu^2) = \mu^2 \Delta f \sim \frac{b+f\mu^2}{m} \Delta m.
\end{equation}
The uncertainty $\Delta m$ in the slope will be roughly (noise in $\delta_i$)/(span of observed $b$ values), or $(\Delta m)^2 \sim [n{\rm Var}(b)]^{-1}$, since the Poisson variance in $\delta_i$ is $1/n$.  Also we note that $\langle m^2 \rangle = P$ since $\delta$ is the slope of the line.  Putting these together,
\begin{equation}
\mu^4 (\Delta f)^2 \sim \frac{1}{nP}\frac{(b+f\mu^2)^2}{{\rm Var}(b)},
\end{equation}
which is very similar to (\ref{sigmaf}).  We see that a small value of ${\rm Var}(b)$ in the target population will lead to a large lever-arm in determining the $b$-intercept that defines $f$, hence a narrow range of biases will produce poor constraints on $f$.

Equation~(\ref{sigmaf}) also exhibits the expected characteristic of a sample-variance-free measurement, namely that $\sigma_f$ drops without bound as the measurement noise is driven to zero $(n\rightarrow\infty)$.  We will see, however, that this gain is not realized, because a real halo population has ${\rm Var}(b)\rightarrow 0$ as we seek large $n$ by going to ever-lower halo masses.

The same single-mode analysis shows that the limit $nP\rightarrow\infty$ yields an uncertainty in the power spectrum amplitude $\sigma_{\ln G}\rightarrow 1/\sqrt{2}$, {\it i.e.} a sample variance limit equal to that for $\ln fG$ in the MS method.

\section{Results}
\begin{figure*}
\resizebox{\hsize}{!}{
\includegraphics[angle=0]{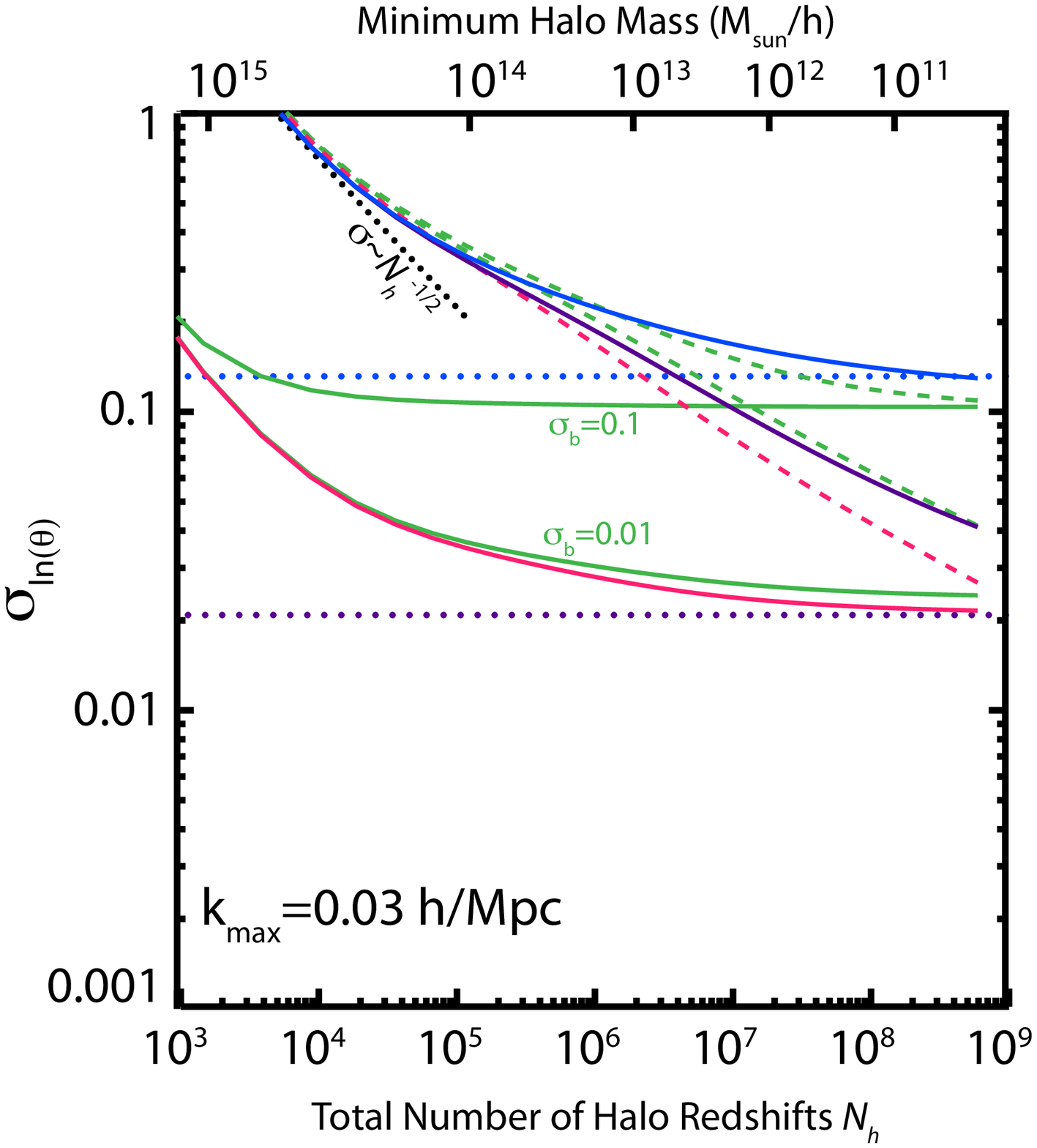}
\includegraphics[angle=0]{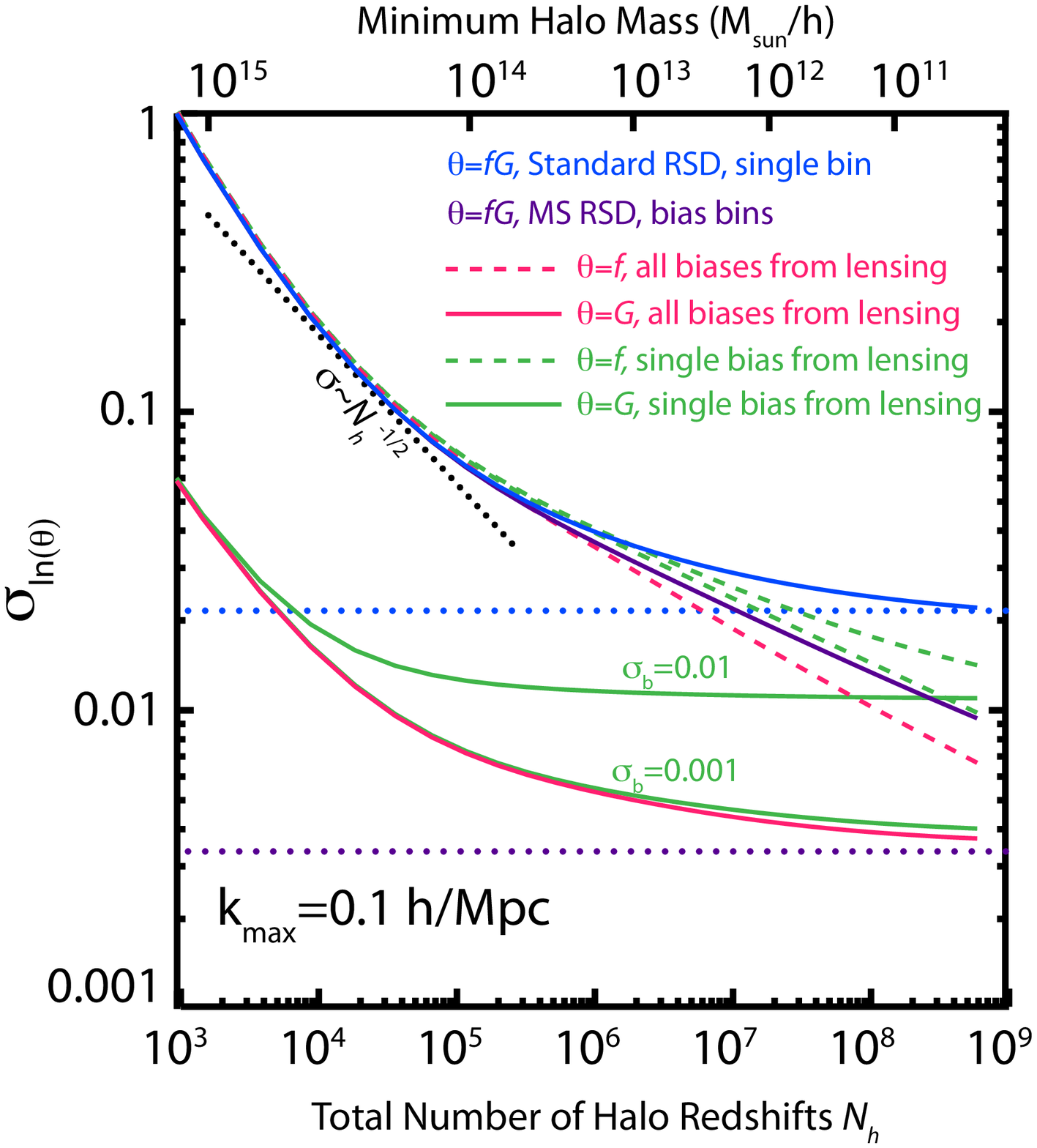}}
\caption[]{
Forecasted uncertainty in growth parameters are plotted against the number of halo redshifts obtained in our fiducial survey: $z=0.5$, $\Delta z=0.1$, $f_{\rm sky}=0.5$.  The survey is assumed to reach all halos above a minimum mass, marked on the top axis.  Left and right panels assume different wavenumbers $k_{\rm max}$ to which linear perturbation theory is sufficiently accurate.  The legend marks the different types of analyses used to extract redshift-space distortion information:  blue and purple curves use single-bin or McDonald-Seljak bias binning to analyze pure galaxy-redshift information and extract the degenerate combination $fG$.  With weak-lensing enables measurements of galaxy bias, $f$ and $G$ can be measured separately to the plotted accuracy, for perfectly known biases (red) and for a single weighted bias uncertain to the marked levels (green).  Combination of weak lensing data with the galaxy redshift survey not only enables a direct measure of $f$, but a distinct measure of $G$ with substantially better precision.  The upper horizontal dotted lines mark the cosmic-variance limits for the standard RSD measure, and the lower dotted lines are the sample variance limits for $\ln fG$ (MS method) or $\ln G$ (lensing$+$redshift methods).
}
\label{sigtheta}
\end{figure*}

Figures~\ref{sigtheta} plots the uncertainty in the growth measures vs the minimum mass $M_{\rm min}$ of halo included in a fiducial redshift survey.  We make the following assumptions for the fiducial survey:
\begin{itemize}
\item The survey covers $f_{\rm sky}=0.5$ of the celestial sphere and a redshift range $\Delta z = 0.1$ centered at $z=0.5$.  In the fiducial $\Lambda$CDM cosmology, the survey volume is 2.6 $h^{-3}{\rm Gpc}^3$, and $f=0.72$ at the survey redshift.
\item The survey obtains a center-of-mass redshift for all halos with $M>M_{\rm min}$. The number and bias of halos vs mass are taken from a Sheth-Tormen mass function \citep{Sheth99}, with model parameters listed therein.
\item The matter fluctuations have a power spectrum given by linear perturbation theory, with the transfer function computed from $CMBFAST$ \citep{Seljak96}.
\item The halos populate the matter distribution by a biased Poisson process.
\item The halos follow the LPT matter velocity field without additional bias or stochasticity, so the Kaiser formulae hold.
\item Only modes with $k<0.1h\,{\rm Mpc}^{-1}$ (optimistic) or $k<0.03h\,{\rm Mpc}^{-1}$ (pessimistic) have theoretical redshift-space predictions of sufficient accuracy to be useful for percent-level constraints of growth.  $N$-body simulations show 20\% departures from the Kaiser formula for redshift-space distortions of halos at $k<0.1h\,{\rm Mpc}^{-1}$ \citep{OkumuraJing}, with significant dependence on halo mass. Substantial advances in theoretical understanding of nonlinear redshift-space effects will be needed to extract percent-level growth constraints at these scales \citep[e.g.][]{Jennings11}.
\end{itemize}
It is desirable to measure the growth function to percent-level accuracy in several redshift bins of this width at $z<1$ in order to test general relativity (in combination with percent-level constraints on $a(t)$ from distance-measurement methods). We note that the total number of independent Fourier modes in the fiducial survey volume is only $N_m=1150$ (43,000) for $k_{\rm max}=0.03h\,{\rm Mpc}^{-1}$ (0.1).  We recall that the standard RSD method has a sample-variance floor of $\sigma_{\ln fG}=\sqrt{21/N_m}$ and therefore a 1\% measure of $fG$ is not attainable for $k_{\rm max}\le0.1h\,{\rm Mpc}^{-1}$.  The sample variance limit of the MS method is $\sigma_{\ln fG}=\sqrt{1/2N_m}$ and the 1\% meaurement is potentially achievable in the optimistic case.
 
\subsection{Surveys without lensing}
Figures~\ref{sigtheta} plot the uncertainty in $\ln fG$ for the Standard RSD and MS methods, and then plot $\sigma_{\ln f}$ and $\sigma_{\ln G}$ independently for the Fixed Bias and the Single Bias Measurement cases.  In this last case, we plot results from two levels of uncertainty on the weighted bias $\bar b$.  Two horizontal axes plot the $M_{\rm min}$ for the halo survey, and the total number of halos in the volume, {\it i.e.} the number of redshift measurements required for the survey.

The Standard RSD analysis (in blue) approaches a sample-variance-limited plateau as expected, with very little improvement once $N_H>10^5$--$10^6$ halo redshifts are obtained.  

The MS analysis (purple) shows significant improvement over the Standard RSD analysis when $>10^6$ redshifts are obtained, or halos $<10^{13}h^{-1} M_\odot$.  The cosmic variance limit for the MS analysis is $\sigma_{\ln fG} > 1/\sqrt{2N_m}$ (dotted purple line) but this is not attained even with surveys of $>10^9$ redshifts.  This is attributable to the limited number of halos with $b>1$, as explained below.  In the conservative case of $k_{\rm max}=0.03h\,{\rm Mpc}^{-1}$, uncertainties of $<10\%$ in $fG$ require $N_h>10^7$ redshifts to be measured in this bin for the MS method.  If the theory supports $k_{\rm max}=0.1h\,{\rm Mpc}^{-1}$, then only $N_h<10^5$ redshifts are needed to reach 10\% accuracy, and $10^7$ redshifts yield a 2\% measure of $fG$.

Uncertainties on $fG$ scale roughly as $N_h^{-1/2}$ when using the standard or MS methods at $N_h<10^5$. In practice a survey has a trade between depth and sky coverage, and one can ask whether better cosmological constraints result from a large, sparse redshift survey or a deeper, smaller-area survey given a fixed survey cost (duration). If every redshift can be attained at equal cost, then this scaling implies a flat trade, i.e. constraint on $fG$ are independent of survey area when $N_h<10^5$.  However the $\sigma$-vs-$N_h$ curve flattens at higher $N_h$, which indicates that one should favor area over depth when $N_h>10^5$.  This would be even more true if redshifts are more expensive to obtain in lower-mass halos.

\subsection{Surveys with lensing data}
The red lines show the uncertainties on $\ln G$ (solid) and $\ln f$ (dashed) in the Fixed Bias case of perfect lensing calibration of all galaxy biases.  We note first that constraints on $\ln f$ (dashed red) are obtained at accuracy similar to the $\ln fG$ constraints with the MS method at equal $N_h$.  Next we note
that the $\ln G$ constraint (solid red) approaches a sample-variance limited plateau much more rapidly than does the MS method.  For $k_{\rm max}=0.03h\,{\rm Mpc}^{-1}$, one can obtain 3\% errors in $\ln G$ with a survey of only $\sim10^5$ halos at $M>10^{14}h^{-1}M_\odot$, $10^4$ times fewer redshifts than are needed for similar $fG$ constraints using the MS method alone, and completely impossible with Standard RSD methods.  This is attributable to the lensing information breaking the degeneracy between bias and matter power spectrum amplitude.

The dashed red line shows that the constraint on $f$ does indeed continue to decrease without apparent bound as $N_h$ increases and shot noise is reduced.  It is disappointing, however, that $\sigma_{\ln f}$ is larger than $\sigma_{\ln G}$ until the survey exceeds $10^9$ redshifts in this bin.  Using Equation~(\ref{sigmaf}) as a guide to the behavior of $\sigma_{\ln f}$, we can see that the survey can attain very large $nP$ but still have the $f$ constraint degraded by an unfavorable lever arm factor $\langle b+f\mu^2\rangle / \sqrt{{\rm Var}(b)}$.  While massive halos have $b>1$ and a significant range of bias, all of the halos with masses $10^{10}$--$10^{12}h^{-1}M_\odot$ have $b$ within a few percent of unity.  As low-mass halos dominate the full sample, ${\rm Var}(b)$ drops, degrading the gains from decreased shot noise.  The scaling is roughly $\sigma_{\ln f}\propto N_h^{-1/3}$.

The green lines show the effect of degrading perfect knowledge of all biases to finite errors on a single weighted bias combination.  We find that the measure of a single weighted galaxy bias results in growth constraints that are nearly as good as the Fixed Bias case.  The degradation in accuracy of parameter $\theta\in\{f,G\}$ is insignificant as long as $\sigma_b \ll \sigma_{\ln \theta}$.  In particular, this means that the weighted bias should be measured to 0.01 (0.001) accuracy for $k_{\rm max}=0.03$ (0.1).  This is not a trivial requirement---a future paper will detail the attainable accuracies in $\bar b$ from the lensing experiments.

\subsection{Modified gravity constraints}
\begin{figure*}
\resizebox{\hsize}{!}{
\includegraphics[angle=0]{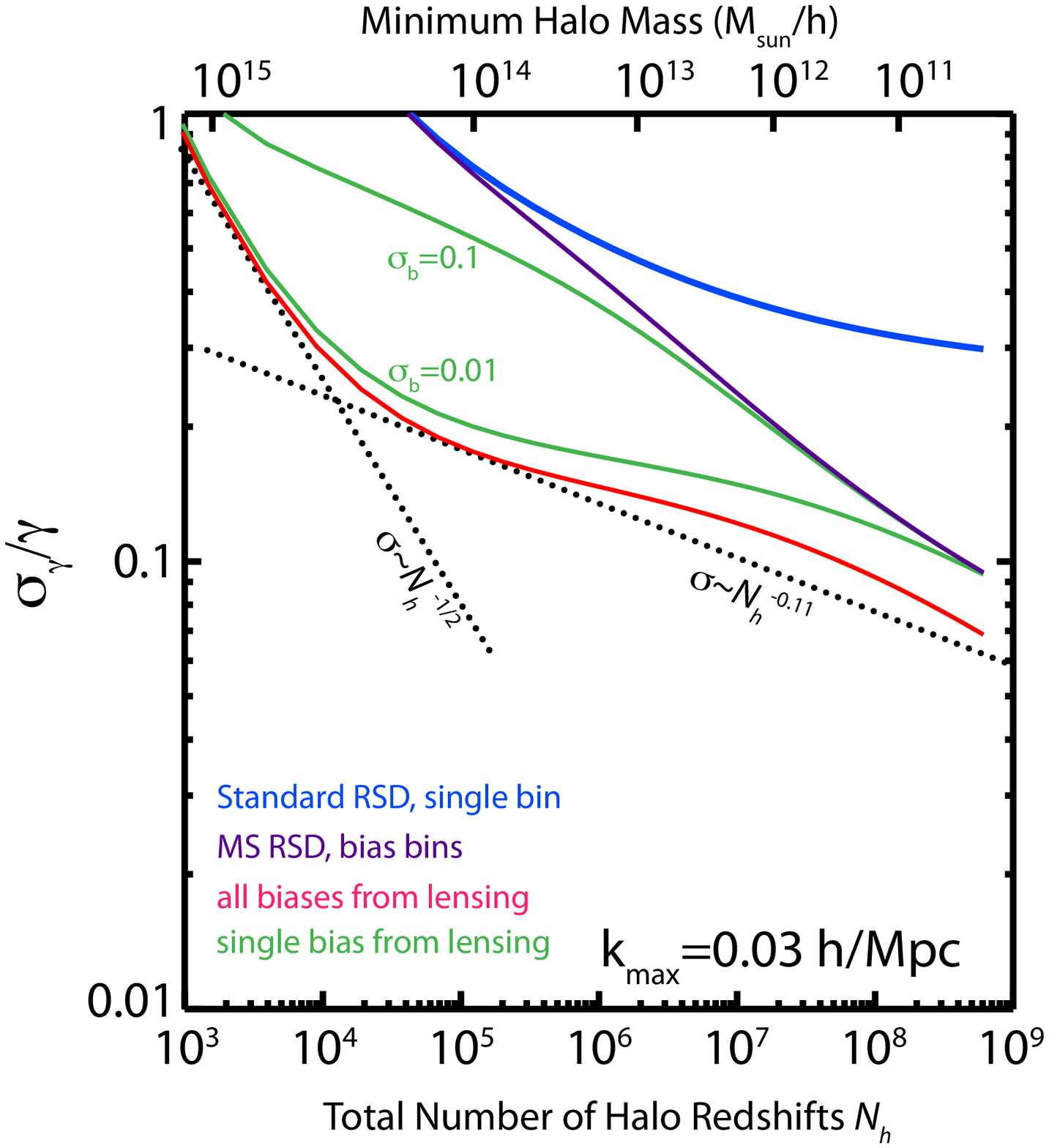}
\includegraphics[angle=0]{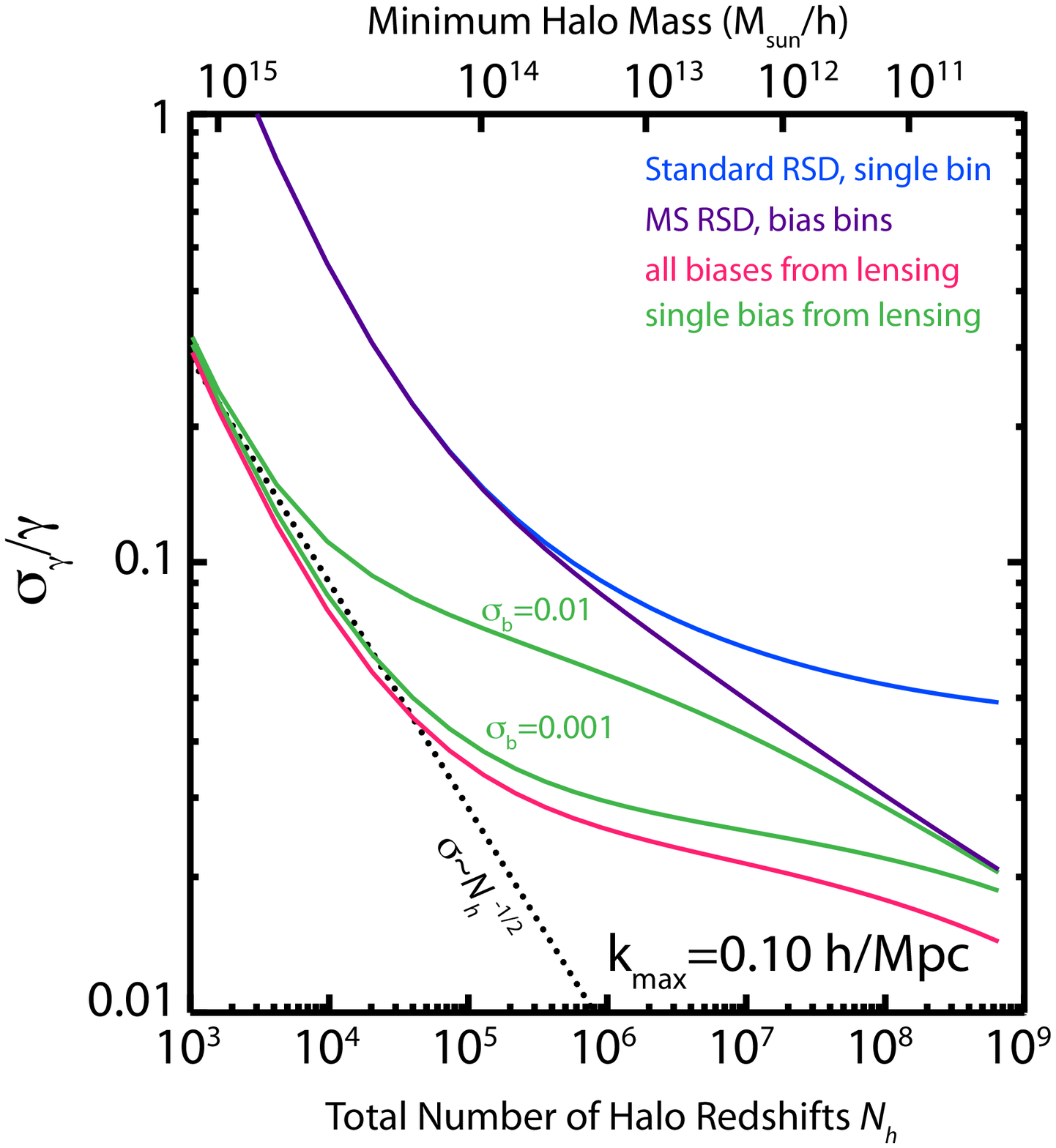}}
\caption[]{
Forecasted constraints on deviations of the growth parameter $\gamma$ in Equation~(\ref{fgamma}) from the General Relativity value of $\gamma=0.55$.  The axes and line colors apply to the same fiducial survey and types of analyses as in Figure~\ref{sigtheta}.  In the case of $k_{\rm max}=0.03h\,{\rm Mpc}^{-1}$ (0.1), a lensing-based bias measurement to accuracy 0.01 (0.001) provides improves constraints on $\gamma$ over a standard redshift-space-distortion experiment by an amount equivalent to a 10-fold increase in survey size.
}
\label{sigmagamma}
\end{figure*}
How do we compare the merits of constraints of $f$ to constraints on $G$ or on $fG$?  For the lensing$+$galaxy surveys, does the correlation between measures of $f$ and $G$ enhance or degrade the detection of modifcations to General Relativity (GR)?  We address both questions by quantifying the constraints on a simplistic model for deviations from GR. \citet[and references therein]{Peebles80,Lahav91,LinderCahn} note that the growth equation for GR is very well approximated by
\begin{equation}
f = \Omega_m^\gamma
\label{fgamma}
\end{equation}
with $\gamma$ in a narrow range near 0.55 for a wide variety of dark-energy models. Gravity theories other than GR tend to have a very different value of $\gamma$. Therefore, a constraint on $\gamma$ may provides sensitive test of GR.

A different choice of $\gamma$ with fixed $\Lambda$CDM expansion history would alter both $G$ and $f$ in a predictable way.  We find numerically that $d(\ln f)/d\gamma\approx 3\, d(\ln G)/d\gamma$ in this model, so we can generically expect constraints on the differential growth $f$ to be $3\times$ more valuable for detecting deviations from GR than constraints on the integrated growth $G$ and $3/4$ as valuable as $fG$ constraints of equal fractional accuracy.
Figures~\ref{sigmagamma} plot the uncertainty on $\gamma$ that would be implied by the forecasted $f$ and $G$ measurements from our fiducial surveys.

The plots suggest that a coincident weak lensing survey greatly enhances the ability of redshift surveys to constrain deviations from GR: the lensing determination of a weighted galaxy bias to 0.01 (0.001) accuracy yields a $\sigma_\gamma$ value that is at least $\sqrt{10}\times$ better than the standard RSD constraints for a survey of fixed $k_{\rm max}=0.03h\,{\rm Mpc}^{-1}$ (0.1) and fixed number of redshifts. In other words, adding a coincident lensing survey to a standard RSD survey improves its growth constraints by an amount equivalent to a $10\times$ expansion of the sky coverage of the RSD survey.
A redshift-only survey using the MS technique can recover some of the advantage of the combined survey, but rather slowly: the MS approach requires $10^8$--$10^9$ redshift measurements to attain the $\gamma$ precision that a combined survey reaches with $10^6$ redshifts.

The scaling of $\sigma_\gamma$ with survey depth is instructive.  For shallow surveys ($M_{\rm min}>10^{14}h^{-1}M_{\odot}$, $N_h=10^4$--$10^5$), the combined survey gains precision quickly, roughly as $N_h^{-1/2}$, the slope at which the depth-vs-area tradeoff will be rather flat.  The gains in precision for deeper combined surveys are quite shallow, $\sigma_\gamma\propto N_h^{-0.1}$.  This reduced return on investment is traceable to the small variation in bias for lower-mass halos. Note that the inflection occurs for a survey that requires redshifts only for galaxy-cluster-mass halos, which are easily identified with imaging surveys at $z<1$.  If biases of such halos can be determined to 0.01 (0.001) fractional accuracy from a coincident lensing survey, then uncertainties in $\gamma$ of $\approx 0.1$ (0.015) per $\Delta z=0.1$ bin can be obtained with only $\approx 10^5$ redshift measurements per bin. Both results would require $>10^8$ redshifts per bin if using the MS technique without lensing, and are better than would be possible from {\em any} standard RSD survey, regardless of depth.

\section{Discussion}
The combination of bias-modulated redshift-space measurements with weak gravitational lensing offers, in principle, a path to unbounded precision in the growth rate $f$ from a survey of finite volume in the Universe.   In addition such techniques allow determination of $f$ without any degeneracy with galaxy bias or the power spectrum, {\it i.e.} one can measure $f$, not just $f/b$ or $f^2P$.  When the redshift survey makes use of dark-matter halos (or the galaxies within them), a practical survey sees very shallow gains in precision on $f$ with increased redshift-survey target density, because only relatively rare halos have the substantial range in bias that is necessary to realize large gains from the MS bias-modulation technique.  Nonetheless we show that the addition of lensing information to redshift surveys yields very large improvements in precision on growth parameters of all kinds.  Typically the precision available from the combined survey technique is equivalent to a 10-fold increase in volume of a redshift-only survey. In other words, the constraints available from the combined survey would require 10 observable Universes without the lensing information on biases.

This significant improvement increases the potential for growth tests at low ($z<1$) redshifts since concerns about the small observable volume are less important.
 Low-redshift galaxy surveys have several advantages, the most obvious being higher fluxes hence less expensive redshift surveys, particularly if we have techniques that benefit from $nP\gg1$.  Another advantage to $z<1$ redshift surveys is that 
one can image and resolve a large number of galaxies that are lensed by the spectroscopic targets, {\it i.e.} are in the background.  At $z<1$, massive halos are helpfully marked by well-developed red sequences.  Our naive expectation is that the most blatant violations of GR should be found in the acceleration epoch.
And the techniques herein ameliorate the greatest disadvantage of $z<1$ surveys, which is the limited observable volume.

Several issues must be addressed before the growth tests forecast herein could be realized.  First, the Kaiser formulae do not hold to percent accuracy in $N$-body tests and improved predictions for non-linear behavior in redshift space are needed---this is true for all the redshift-space distortion methods, not just the combined lensing$+$redshift surveys.  Particularly crucial for the combined method is that the covariance in Equation~(\ref{cij}) between galaxy bins be accurate, namely that the velocity field is unbiased and the real-space covariance between galaxy bins be the product of each bin's covariance with the mass.  Other issues we will address in future work are the impact of projections and shape noise on the determination of bias via weak lensing.  The optimal depth for the redshift survey is also an issue: Figure~\ref{sigmagamma} indicates limited gains for surveys of halos $<10^{14}h^{-1}M_\odot$, but it is possible that the determination of bias by lensing-galaxy cross-correlation will require a deeper galaxy sample in order to reach low levels of stochasticity and exploit the absence of sample variance.

This work is supported by grant AST-0908027 from the NSF and DOE grant
DE-FG02-95ER40893. 

\bibliography{nocv}

\begin{thebibliography}{}

\bibitem[\protect\citeauthoryear{{Albrecht}, {Amendola}, {Bernstein}, {Clowe},
  {Eisenstein}, {Guzzo}, {Hirata}, {Huterer}, {Kirshner}, {Kolb} \&
  {Nichol}}{{Albrecht} et~al.}{2009}]{FoMSWG}
{Albrecht} A.,  {Amendola} L.,  {Bernstein} G.,  {Clowe} D.,  {Eisenstein} D.,
  {Guzzo} L.,  {Hirata} C.,  {Huterer} D.,  {Kirshner} R.,  {Kolb} E.,
  {Nichol} R.,  2009, arXiv:0901.0721

\bibitem[\protect\citeauthoryear{{Bernstein} \& {Jain}}{{Bernstein} \&
  {Jain}}{2004}]{BJ04}
{Bernstein} G.,  {Jain} B.,  2004, \apj, 600, 17

\bibitem[\protect\citeauthoryear{{Blake} \& {et al.}}{{Blake} \& {et
  al.}}{2011}]{Blake}
{Blake} C.,  {et al.} 2011, arxiv:1104.2948

\bibitem[\protect\citeauthoryear{{Cai}, {Bernstein} \& {Sheth}}{{Cai}
  et~al.}{2011}]{CBS11}
{Cai} Y.,  {Bernstein} G.,    {Sheth} R.~K.,  2011, \mnras, 412, 995

\bibitem[\protect\citeauthoryear{{Gil-Mar{\'{\i}}n}, {Wagner}, {Verde},
  {Jimenez} \& {Heavens}}{{Gil-Mar{\'{\i}}n} et~al.}{2010}]{GM10}
{Gil-Mar{\'{\i}}n} H.,  {Wagner} C.,  {Verde} L.,  {Jimenez} R.,    {Heavens}
  A.~F.,  2010, \mnras, 407, 772

\bibitem[\protect\citeauthoryear{{Hamaus}, {Seljak}, {Desjacques}, {Smith} \&
  {Baldauf}}{{Hamaus} et~al.}{2010}]{Hamaus10}
{Hamaus} N.,  {Seljak} U.,  {Desjacques} V.,  {Smith} R.~E.,    {Baldauf} T.,
  2010, \prd, 82, 043515

\bibitem[\protect\citeauthoryear{{Jennings}, {Baugh} \& {Pascoli}}{{Jennings}
  et~al.}{2011}]{Jennings11}
{Jennings} E.,  {Baugh} C.~M.,    {Pascoli} S.,  2011, \mnras, 410, 2081

\bibitem[\protect\citeauthoryear{{Kaiser}}{{Kaiser}}{1987}]{Kaiser}
{Kaiser} N.,  1987, \mnras, 227, 1

\bibitem[\protect\citeauthoryear{{Lahav}, {Lilje}, {Primack} \& {Rees}}{{Lahav}
  et~al.}{1991}]{Lahav91}
{Lahav} O.,  {Lilje} P.~B.,  {Primack} J.~R.,    {Rees} M.~J.,  1991, \mnras,
  251, 128

\bibitem[\protect\citeauthoryear{{Linder} \& {Cahn}}{{Linder} \&
  {Cahn}}{2007}]{LinderCahn}
{Linder} E.~V.,  {Cahn} R.~N.,  2007, Astroparticle Physics, 28, 481

\bibitem[\protect\citeauthoryear{{McDonald} \& {Seljak}}{{McDonald} \&
  {Seljak}}{2009}]{MS}
{McDonald} P.,  {Seljak} U.,  2009, \jcap, 10, 7

\bibitem[\protect\citeauthoryear{{Okumura} \& {Jing}}{{Okumura} \&
  {Jing}}{2011}]{OkumuraJing}
{Okumura} T.,  {Jing} Y.~P.,  2011, \apj, 726, 5

\bibitem[\protect\citeauthoryear{{Peebles}}{{Peebles}}{1980}]{Peebles80}
{Peebles} P.~J.~E.,  1980, {The large-scale structure of the universe}

\bibitem[\protect\citeauthoryear{{Pen}}{{Pen}}{2004}]{Pen04}
{Pen} U.,  2004, \mnras, 350, 1445

\bibitem[\protect\citeauthoryear{{Seljak} \& {Zaldarriaga}}{{Seljak} \&
  {Zaldarriaga}}{1996}]{Seljak96}
{Seljak} U.,  {Zaldarriaga} M.,  1996, \apj, 469, 437

\bibitem[\protect\citeauthoryear{{Sheth} \& {Tormen}}{{Sheth} \&
  {Tormen}}{1999}]{Sheth99}
{Sheth} R.~K.,  {Tormen} G.,  1999, \mnras, 308, 119

\bibitem[\protect\citeauthoryear{{Takada} \& {Jain}}{{Takada} \&
  {Jain}}{2004}]{TakadaJain}
{Takada} M.,  {Jain} B.,  2004, \mnras, 348, 897

\bibitem[\protect\citeauthoryear{{Tegmark}, {Taylor} \& {Heavens}}{{Tegmark}
  et~al.}{1997}]{TTH}
{Tegmark} M.,  {Taylor} A.~N.,    {Heavens} A.~F.,  1997, \apj, 480, 22

\bibitem[\protect\citeauthoryear{{White}, {Song} \& {Percival}}{{White}
  et~al.}{2009}]{WSP09}
{White} M.,  {Song} Y.,    {Percival} W.~J.,  2009, \mnras, 397, 1348

\end{thebibliography}
\bibliographystyle{mn2e}

\end{document}